\documentclass[12pt]{article}
\usepackage{amsfonts,amssymb,amsmath,graphicx,epic,graphics,color}

\newcommand{\N}{{\cal N}}

\def\thefootnote{\fnsymbol{footnote}}

\begin{document}

\thispagestyle{empty}
\renewcommand{\thefootnote}{\fnsymbol{footnote}}

{\hfill \parbox{3.1cm}{\mbox{}
         \\
}}

\bigskip\bigskip

\begin{center} \noindent \Large \bf
Worldsheet correlators in AdS$_3$/CFT$_2$\\
\end{center}

\bigskip\bigskip\bigskip

\setcounter{footnote}{2}

\begin{large}
\centerline{Matthias R.\ 
Gaberdiel\footnote{Email: {\tt gaberdiel@itp.phys.ethz.ch}} 
and 
Ingo Kirsch\footnote{Email: {\tt kirsch@phys.ethz.ch}}}
\end{large}

\begin{center}

\textit{Institut f\"ur Theoretische Physik, ETH Z\"urich \\ 
CH-8093 Z\"urich, Switzerland}\\
\end{center}

\bigskip\bigskip
\bigskip\bigskip

\renewcommand{\thefootnote}{\arabic{footnote}}
\setcounter{footnote}{0}

\centerline{\bf \small Abstract}
\medskip

{\small The AdS$_3$/CFT$_2$ correspondence is checked beyond the
supergravity approximation by comparing correlation functions.
To this end we calculate 2- and 3-point functions on the sphere of 
certain chiral primary operators for strings on 
$AdS_3 \times  S^3 \times T^4$. These results 
are then compared with  the corresponding amplitudes in the dual
2-dimensional conformal field theory. In the limit of small string
coupling, where the sphere diagrams dominate the string perturbation
series, beautiful agreement is found.}

\section{Introduction}

Over the last few years, increasing evidence has been accumulated that
(in the large $N$ limit) $4$-dimensional gauge theories have a dual
description in terms of a higher-dimensional string theory. This
so-called AdS/CFT correspondence \cite{Maldacena97,Gubser98,Witten98}
has predominantly been tested in the supergravity approximation of
string theory at small curvatures (see however the analysis in the
plane-wave limit \cite{BMN}).  On the other hand, the AdS/CFT
correspondence actually relates the weak coupling regime of gauge
theory to backgrounds with high curvature in string theory. It would
therefore be very interesting to understand and check the
correspondence for such configurations. Unfortunately, the high
curvature regime of type IIB string theory on $AdS_5 \times S^5$ that
is dual to four-dimensional ${\cal N}=4$ $SU(N)$ SYM theory is
difficult to control at present.

The situation is different for the AdS$_3$/CFT$_2$ duality for which
both the ten-dimen\-sio\-nal string theory as well as the
two-dimensional boundary conformal field theory are explicitly known
(for a review see \cite{David}).  Starting from the intersection of
$N_1$ D1-branes and $N_5$ D5-branes compactified on a four-torus
$T^4$, one finds $AdS_3 \times S^3 \times T^4$ as the corresponding
near-horizon geometry. After S-duality, the worldsheet theory with
target space $AdS_3 \times S^3 \times T^4$ is an $\N=1$ $SL(2) \times
SU(2)$ WZW model which is in principle solvable to all orders in
$\alpha'$. The conjectured dual field theory arises as the infrared
fixed point theory living on the D1-D5 system.  The fixed point can be
described by an $\N=(4,4)$ sigma model whose target space is a
deformation of the symmetric product orbifold ${\rm Sym}(\tilde
T^4)^{N}=(\tilde T^4)^{N}/S_N$, where $\tilde T^4$ is related to the
four-torus $T^4$ \cite{Giveon} and $S_N$ is the permutation group of
$N$ symbols.  It has been argued in \cite{Larsen} that the orbifold
limit corresponds to the point $(N_1, N_5)$=$(N, 1)$ in the D1-D5
moduli space.

In this paper, we will subject this correspondence to a non-trivial
test by comparing the correlators of certain chiral primary fields in
both theories. We will follow the approach outlined in
\cite{Maldacena} that explains how the correlators of the boundary
conformal field theory can be obtained from the world-sheet
correlators of the string theory.\footnote{For correlation functions
  involving more than three fields, the analysis of Gopakumar
  \cite{Gopakumar} should be relevant.}  There it is suggested that
the integration of vertex operators over the worldsheet
yields the corresponding boundary operator. This is motivated by the
fact that the continuous $SL(2)$ representation labels $x,\bar x$ are
identified with the complex coordinates in the boundary conformal
field theory, as already noted in \cite{deBoer}.

Here we shall apply this programme to the full theory, including the
$SU(2)$ WZW model as well as the theory on $T^4$.  We shall
concentrate on the chiral primary operators that correspond to the
$n$-cycle twist operators in the boundary conformal field
theory~\cite{Kutasov}. In the boundary conformal field theory their
correlation functions were first calculated for a special case  
in \cite{Jevicki} and then, in general, in \cite{Lunin,Lunin2} 
(see also \cite{Arutyunov2} for earlier work). The correlation
functions in the  world-sheet theory (on the sphere) can be reduced to
standard 3-point functions of the $SU(2)$ and $SL(2)$ WZW models that
have been determined before in \cite{Zamolodchikov, Dotsenko} and 
\cite{Teschner1997, Teschner1999, FZZ}, respectively. The comparison
of the 2-point functions determines the relative normalisation
constants between the field operators on both sides of the
correspondence.  The comparison of the 3-point functions is then a
non-trivial consistency check. In the large $N$ limit (in which the
dominant string contribution comes from the sphere diagram) we find
beautiful agreement.  \smallskip

In the supergravity approximation, the correlators of these chiral
primary fields were previously computed in
\cite{Mihailescu,Arutyunov}. A quantitative comparison with the
boundary conformal field theory  at the orbifold point (for which the
conformal field theory correlators are known) is however impeded by
the fact that the corresponding D1-D5 system is at strong string
coupling. By S-duality this can be mapped to the F1-NS5 system at
small string coupling, but then the $AdS_3$ radius is of order of the
string length, and hence $\alpha'$ corrections have to be taken into
account.
\medskip

The paper is organised as follows. In section~\ref{sec2}, we review
the symmetric orbifold theory and the correlators of $n$-cycle twist
operators found by Lunin and Mathur \cite{Lunin,Lunin2}. In
section~\ref{sec3}, we then compute correlation functions of the
corresponding primary vertex operator in the worldsheet theory with
target space  $AdS_3 \times S^3 \times T^4$ and compare them with
those in the boundary conformal field theory. Our conclusions are
contained in section~4. There are three appendices where some of the
more technical material is explained.

%
\section{Sigma model on the symmetric orbifold} \label{sec2}
%

The Higgs branch of the D1-D5 system compactified on $T^4$ flows in
the infrared to a two-dimen\-sional $(4,4)$ superconformal field
theory with conformal charge $c=6N_1 N_5$. The global part of the
$\N=(4, 4)$ superconformal algebra forms the supergroup $SU(1,1|2)_L
\times SU(1,1|2)_R$ and contains the R-symmetry group $SU(2)_L \times
SU(2)_R$. It is generally believed that the target space of this
particular $(4,4)$ SCFT is identical to the instanton moduli space
${\cal P}$ of $N_1$ instantons in the $U(N_5)$ gauge theory on $T^4$.
This idea is motivated by the fact that the D1-branes can be viewed as
instantons of the low-energy theory of the D5-branes. The appropriate
low-energy description of the D1-D5 system is therefore given by a
sigma model with target space ${\cal P}$.

\subsection{The $n$-cycle twist operators}

The moduli space ${\cal P}$ of the infrared $(4,4)$ SCFT is the
symmetric product orbifold ${\rm Sym}(\tilde T^4)^{N}=(\tilde
T^4)^{N}/S_N$, where $S_N$ denotes the permutation group of $N=N_1N_5$
symbols.  The four-torus $\tilde T^4$ is closely related to the $T^4$
in the worldsheet model, for the exact relation see
Ref.~\cite{Giveon}. The moduli space is parametrized by the scalars
$X_A^i$, where the $A=1,...,N$ runs over the $N$ copies of $\tilde
T^4$, and $i=1,2,3,4$ denotes the vector index in $\tilde T^4$.  The
orbifold action on $(\tilde T^4)^N$ by the symmetric group $S_N$ means
that any point $(X_1, ...,X_N)$ on $(\tilde T^4)^N$ is identified with
the points obtained by any permutation of the $X_A$ ($A=1,...,N$).

The orbifold theory consists of the invariant operators of the
original theory, together with operators from the twisted sectors. For 
non-abelian orbifolds the twisted sectors are associated to the
conjugacy classes of the orbifold group, which is $S_N$ in our
case. These conjugacy classes are labelled by partitions of $N$ into
positive integers, 
\begin{align}
\sum_{l=1}^N l k_l = N = N_1N_5 \ ,
\end{align}
corresponding to the permutations with $k_l$ cycles of length $l$. We
are mainly interested in the conjugacy class with one cycle of length
$n$, {\it i.e.}\ $k_n=1$ and $k_1=N-n$. The corresponding permutations
are of the form
\begin{align}
(X_{A_1} \rightarrow X_{A_2}, ..., X_{A_n} \rightarrow X_{A_1})\ ,
\qquad X_{B} \rightarrow X_{B} \ , \quad 
B \not\in \{A_1,\ldots, A_n\} \ ,
\end{align}
where $A_1 \neq A_2 \neq ...\neq A_n \in \{1,...,N\}$. We call the
corresponding operators $n$-cycle twist operators and denote them by 
\begin{align} \label{twistoperator}
\Sigma^{(n)} (x,\bar x) \,, \qquad
j'=\tfrac{n-1}{2} = 
0,\tfrac{1}2, 1, ...,\tfrac{N-1}{2} \ .
\end{align}  
The label $j'=\bar{j}'=(n-1)/2$ denotes the charge under the $SU(2)_L
\times SU(2)_R$ R-symmetry group, and their conformal dimension is
$h=\bar h=(n-1)/2$. In particular, they therefore satisfy $h=j'$, and
thus define chiral primary fields. These fields are the analogues of
the single-trace operators in the $\N=4$ super Yang-Mills theory. As
such, they correspond to single-particle states. The theory obviously
also has operators for conjugacy classes with more than one cycle;
these describe multi-particle states (see also \cite{Lunin,
  Lunin2,David}).  In the following we shall however restrict
attention to the above $n$-cycle twist operators.

\subsection{Correlators of $n$-cycle twist operators}

The chiral primaries of the $(4, 4)$ superconformal field theory
actually fall into short multiplets of the supergroup $SU(1,1 | 2)_L
\times SU(1,1|2)_R$. In particular, they therefore define
representations of the $SU(2)_L\times SU(2)_R$ $R$-symmetry.  The
above operators $\Sigma^{(n)}$ are the highest weight states with
respect to this action; if we include the $m'$ and $\bar{m}'$
dependence we should therefore write them as $\Sigma^{(n)} \equiv
\Sigma^{(n)}_{j',j'}$. More generally, we may thus also define the
operators $\Sigma^{(n)}_{m',\bar m'}$ by acting on 
$\Sigma^{(n)}_{j',j'}$ with the $SU(2)_L$ and $SU(2)_R$ lowering
operators --- see \cite{Lunin} for more details. There are also
different chiral primary operators that can be obtained by multiplying
$\Sigma^{(n)}$ with appropriate combinations of spinors $\psi^i_A$
\cite{David}, the superpartners of the $X^i_A$. However, in this paper
we shall only consider the chiral 
primaries~$\Sigma^{(n)}_{m',\bar m'}$. 

The two- and three-point functions of the chiral twist operator
$\Sigma^{(n)}_{m',\bar m'}$ have been calculated, using path integral
methods, in \cite{Lunin, Lunin2}. The two-point function is given
by~\cite{Lunin}\footnote{The twist operator $\Sigma^{(n)}$ is denoted
  by $O_n^{--}$ in \cite{Lunin} and defined by Eq.~(6.42) therein.}
\begin{align}\label{2ptLunin}
\langle \Sigma^{(n)}_{m',\bar m'}(x_1,\bar x_1) 
\Sigma^{(n)}_{-m',-\bar m'}(x_2,\bar x_2) \rangle
= \frac{1}{|x_{12}|^{4h}} \ ,
\end{align}
where $h=(n-1)/2$. Here we have normalised the fields in the usual way. 
With this normalisation the three-point function is given by
\begin{align} \label{3ptLunin}
&\langle \Sigma^{(n_1)}_{m'_1,\bar m'_1}(x_1,\bar x_1) \,
\Sigma^{(n_2)}_{m'_2,\bar m'_2}(x_2, \bar x_2)  \,
\Sigma^{(n_3)}_{m'_3,\bar m'_3}(x_3, \bar x_3) \rangle = 
\delta^2({\textstyle  \sum_{a=1}^3 m'_a}) C_{n_1,n_2,n_3}  \, 
\prod_{i<j} \frac{1}{|x_{ij}|^{2h_{ij}}} \ ,
\end{align}
with $h_{12}=h_{1}+h_{2}-h_{3}$, {\it etc}, and $\delta^2$ is the
product of the Kronecker $\delta$ for the barred and unbarred 
$\sum_a m_a'$. For the choice $(d \geq 0)$
\begin{align} \label{mprimes}
m'_1=\bar m'_1=j'_1-d\ ,\qquad
m'_2=\bar m'_2=j'_2\ ,\qquad
m'_3=\bar m'_3=-(j'_1+j'_2-d)=-j'_3 \ ,
\end{align}
the fusion coefficients $C_{n_1,n_2,n_3} = C^{(1)}_{n_1,n_2,n_3} \,
C^{(2)}_{n_1,n_2,n_3}$ are given by
\begin{align}
C^{(1)}_{n_1,n_2,n_3} = \frac{s^2 \, d!\,
(n_1-d-1)!}{n_1 \, n_2 \, n_3\, (n_1-1)!}  \ , \quad 
C^{(2)}_{n_1,n_2,n_3}=
\frac{\sqrt{n_1 n_2 n_3 (N-n_1)!(N-n_2)!(N-n_3)!}}{(N-s)! \sqrt{N!}}  
\end{align}
with $2 s= n_1+n_2+n_3-1$ and $2 d = n_1+n_2-n_3-1$. Here
$C^{(1)}_{n_1,n_2,n_3}$ is the coefficient in the three-point function
coming from single representatives of the conjugacy classes; the
factor $C^{(2)}_{n_1,n_2,n_3}$ on the other hand results from the
summation over all elements in the given conjugacy classes. Since
$n_j$ denotes the cycle length we obviously have $n_j\leq N$ for
$j=1,2,3$. In addition, as explained in \cite{Lunin} we also have that
$s\leq N$. For the special case $d=0$, the three-point function
(\ref{3ptLunin}) was first found in \cite{Jevicki}.

For later convenience, we also write the correlators in terms of the
labels $j'=(n-1)/2$, identifying 
$V^{}_{j', m', \bar m'} \equiv \Sigma^{(n)}_{m',\bar m'}$. 
As in Eq.~(\ref{2ptLunin}), we normalise the two-point
function in the boundary conformal field theory as
\begin{align} \label{2ptbCFT}
\langle V^{}_{j',m',\bar m'} (x_1,\bar x_1) \, 
V^{}_{j',-m',-\bar m'} (x_2,\bar x_2) \rangle 
= \frac{1}{|x_{12}|^{4j'}} \ .
\end{align}
The corresponding three-point function then equals  
\begin{align}\label{3ptbCFT}
&\langle V^{}_{j'_1,m'_1,\bar m'_1} (x_1,\bar x_1) \,
V^{}_{j'_2,m'_2,\bar m'_2} ( x_2, \bar x_2) \,
V^{}_{j'_3,m'_3,\bar m'_3} ( x_3, \bar x_3) \rangle 
= 
C^{\rm bcft}_{j'_1,j'_2,j'_3} \prod_{i<j}
  \frac{1}{|x_{ij}|^{2j'_{ij}}} \ ,
\end{align}
where $C^{\rm bcft}_{j'_1,j'_2,j'_3} = C^{(1)}_{j'_1,j'_2,j'_3}
C^{(2)}_{j'_1,j'_2,j'_3}$ is given by ($d = j'_{12} \geq 0$)
\begin{align}
C^{(1)}_{j'_1,j'_2,j'_3} &=  (j'_1+j'_2+j'_3+1)^2 
\left( \frac{\Gamma(j'_{13}+1)\Gamma(j'_{12}+1)}
{\Gamma(2j'_1+1) \prod_{i=1}^3 (2j'_i+1)} \right)  \\
C^{(2)}_{j'_1,j'_2,j'_3} &= \left( \frac{\prod_{i=1}^3 (2j'_i+1)
\Gamma(N-2j'_i)}
{\Gamma(N-j'_1-j'_2-j'_3)^2 \Gamma(N+1)} \right)^{1/2} \ .
\end{align}
Here the $m'$ labels are chosen as in Eq.~(\ref{mprimes})
and $j'_{12}=j'_1+j'_2-j'_3$, {\it etc.}

As before, there are some restrictions on the quantum numbers $j'_i$
($i=1,2,3$). In particular, the bounds $n_i = 2j'_i+1 \leq N$ and 
$s\leq N$ in Eq.~(\ref{3ptLunin}) translate into 
\begin{align} \label{constraintsj}
0 \leq j'_i \leq \textstyle \frac{N-1}{2} 
\,,\qquad j_1'+ j_2'+j_3' \leq N-1 \ .
\end{align}
For the comparison with the string correlators on the sphere only the
large $N$ limit will be relevant; in this limit, the total coefficient
simplifies to 
\begin{align}\label{Nlim}
&C^{\rm bcft}_{j'_1,j'_2,j'_3} 
\stackrel{N \rightarrow \infty}{=}
\frac{(j'_1+j'_2+j'_3+1)^2 }{\sqrt{N}\prod_i (2j'_i+1)^{\frac{1}{2}}} 
 \frac{ \Gamma(j'_{13}+1)\Gamma(j'_{12}+1)}{\Gamma(2j'_1+1)} \ .
\end{align}
In the next section we will reproduce this result by computing the
three-point function (on the sphere) of the dual worldsheet vertex 
operators of string theory on $AdS_3 \times S^3 \times T^4$.

%
\section{Superstring theory on $AdS_3 \times S^3 \times T^4$}
\label{sec3}
%

The AdS/CFT correspondence relates the above 2-dimensional conformal
field theory with string theory on $AdS_3 \times S^3 \times T^4$.
This is the near-horizon geometry of the D1-D5 system. By S-duality we
can relate this to the configuration of fundamental strings and NS5
branes. The latter system then has a description in terms of a WZW
model. More precisely, the relevant world-sheet theory is the product
of an $\N=1$ WZW model on $H_3^+$, an $\N=1$ WZW model on $S^3\cong
SU(2)$ and an $\N=1$ $U(1)^4$ free superconformal field theory.

This WZW model has the affine world-sheet symmetry 
$\widehat{sl}(2)_k \times \widehat{su}(2)_{k'} \times u(1)^4$. 
In the following we shall (as is usual for $\N=1$ WZW models) decouple
the fermions from the currents; the resulting bosonic currents (that
then commute with the fermions) are $J^a$ for $SL(2)$, and
$K^a$ for $SU(2)$. We denote the free fermions that come from the
$SL(2)$ part by $\psi^a$, while those from the $SU(2)$ part are 
$\chi^a$. (In either case, $a$ takes the three values $a=(+,0,-)$.) 
Finally the $u(1)^4$ symmetry is described in terms of free bosons
as $i\partial Y^i$, and the corresponding free fermions are 
$\lambda^i$ ($i=1,2,3,4$).

Criticality of the fermionic string on $AdS_3 \times S^3$ requires the
identification of the levels $k$ and $k'$ \cite{Giveon},
$k=k'$.\footnote{We shall use the convention that $k$ and $k'$ denote
the levels of the supersymmetric $\N=1$ theory; the levels of the
bosonic affine symmetries are then shifted by the dual Coxeter numbers,
$k_{\rm bos} = k_{\rm susy} + 2$ and 
$k'_{\rm bos} = k'_{\rm susy} - 2$.}
Furthermore, the $SU(2)_R \times SU(2)_L$ 
global $R$-symmetry of the boundary conformal field theory corresponds
to the isometry $SO(4)=SU(2)\times SU(2)$ of the three-sphere $S^3$.
In the $SU(2)$ WZW model this symmetry is identified with the
horizontal subalgebras generated by the $K^a_0$ and $\bar{K}^a_0$
of the affine $\widehat{su}(2)_{k'}$ symmetry. The levels $k=k'$ are
to be identified with the number $N_5$ of NS5-branes,
\begin{align}
k=k'= N_5 \ .
\end{align}    
Finally, it is common lore in the ${\rm AdS}_3$/CFT$_2$ correspondence
to interpret the continuous $SL(2)$ representation labels $(x, \bar
x)$ (that will be introduced momentarily) with the complex coordinates
of the boundary conformal field theory \cite{deBoer}.

For the comparison of worldsheet and boundary CFT correlation
functions, we also have to identify the point in the moduli space of
the D1-D5 system (or better the S-dual F1-NS5 system) at which string
theory on $AdS_3 \times S^3 \times T^4$ is dual to the symmetric
orbifold theory. In \cite{Larsen} it was argued that the symmetric
orbifold corresponds to the point $N_1=N$, $N_5=1$, where $N_1$ is the
number of fundamental strings and $N_5$ the number of NS5-branes. At
this point the $AdS_3$ radius $R_{\rm AdS}= l_s (g^2_6 N_1
N_5)^{\frac{1}{4}} = l_s \sqrt{N_5}$ (in the F1-NS5 system) is of
order of the string scale ($g^2_6=N_5/N_1$ is the six-dimensional
string coupling), and supergravity is not a good approximation any
more. We must therefore consider the full worldsheet theory.

On the other hand, we will only be able to calculate the world-sheet
correlators on the sphere, whereas the full string theory amplitude
also involves the contribution from arbitrary genus Riemann
surfaces. We therefore need to work in the limit where $g_s$ is small;
since we have \cite{Giveon, Kutasov}
\begin{align}
g^2_s = \frac{N_5}{N_1} \, {\rm Vol}(T^4) \ ,
\end{align}
$g_s$ is small if ${\rm Vol}(T^4) N_5 \ll N_1$. By T-duality arguments
\cite{Giveon}, the volume can be chosen as ${\rm Vol}(T^4)\geq 1$. At
the point $N_1=N$, $N_5=1$ and fixed volume ${\rm Vol}(T^4)$, the
worldsheet theory is weakly coupled if $N$ is large such that ${\rm
  Vol}(T^4) \ll N$. Note that the AdS space is still strongly curved
in the large $N$ limit and the WZW model is the only reliable
description.

Unfortunately, the worldsheet model is not properly defined for
$(N_1, N_5) = (N,1)$, since the bosonic level $k'_{\rm bos}=N_5-2=-1$ 
is negative at this point. We therefore choose $N_1$ and $N_5$ as  
$(N_1,N_5)=(N/N_5, N_5)$ with $N_5>1$ but fixed, such that $g_s$
remains small for large $N$.  Even though we will compute the
correlators at a point in the moduli space different from the orbifold
point, we  will see below that the worldsheet correlators do not
depend on the actual factorisation of $N=N_1 N_5$, but only on $N$. It
is therefore natural to expect that the results will agree with
those expected from the orbifold theory, at least at large $N$.

\subsection{The chiral primaries in the worldsheet theory}

After these preliminary discussions we now need to study the fields of
this world-sheet theory. First we describe the left-moving fields. The
highest weight states of the $SL(2)$ and $SU(2)$ WZW model are denoted
by $\Phi_{j,m}$ and $\Phi'_{j',m'}$, respectively; their conformal
dimensions are
\begin{align}
\Delta_j=-\frac{j(j-1)}{k} \quad \hbox{and} \quad
\Delta'_{j'}=\frac{j'(j'+1)}{k'} \ .
\end{align}
The OPE's of the bosonic $SL(2)$ currents $J^a$ with the primary
fields $\Phi_{j,m}$ are given as
\begin{align}
J^\pm(z) \Phi_{j,m}(w)&= \frac{m \mp (j-1)}{z-w} \Phi_{j,m\pm 1}(w)\,
\,+ : J^\pm \Phi_{j,m}(w): \,+ \,... \,,\nonumber\\
J^0(z) \Phi_{j,m}(w)&=\frac{m}{z-w} \Phi_{j,m}(w)\,
\,+ : J^0 \Phi_{j,m}(w): \,+ \,... \ , \label{currents}
\end{align}
where $j\geq 1$.\footnote{We are using here the same conventions as in
KLL \cite{Kutasov} except that $j_{\rm KLL}=j-1$. However these  
representations are not of the type discussed by \cite{Maldacena} in
their analysis of the bosonic WZW model corresponding to $SL(2)$.} 
The OPEs of the $K^a$-currents with the primary fields $\Phi'_{j',m'}$
are similar, but we do not need them in the following; more 
details about the $SL(2)$ and $SU(2)$ primary fields and their
correlators can be found in appendix~\ref{AppA}.

To construct the `chiral primaries' of the worldsheet theory on 
$AdS_3 \times S^3 \times T^4$ that correspond to the chiral primaries 
of the boundary conformal field theory described in section~2, we now
need to tensor these left-moving fields with corresponding
right-moving fields. Furthermore, we have to make sure that the fields
survive the GSO-projection and have the correct ghost number. In the
following we shall only consider the NS sector. The left-moving fields
of interest are then \cite{Kutasov} 
\begin{align} \label{wsW}
 {\cal W}_{j',m',m}&= e^{-\phi}(\psi \Phi)_{j-1,m} \Phi'_{j',m'}\ ,
 \qquad(j'=0,\textstyle\frac{1}{2},...,\frac{k'}{2}-1)
\end{align}
where
\begin{align} 
 (\psi \Phi)_{j-1,m} &= \psi^0 \Phi_{j,m} - \frac{1}{2} \psi^+ \Phi_{j,m-1}
 - \frac{1}{2} \psi^- \Phi_{j,m+1}
 \label{abbrev1}\ . 
\end{align}
The bosonised superghost fields $e^{-\phi}$ ensure that the operator 
${\cal W}_{j',m',m}$ has the correct ghost number $-1$. Furthermore, 
in order to guarantee that $ {\cal W}_{j',m',m}$ has also the correct
conformal weight we need to set
\begin{align}
j=j'+1 \ ,\label{jj'}
\end{align}
which justifies dropping the $j$ label in the definition of 
${\cal W}_{j',m',m}$. In fact, using that $k=k'$, we have
\begin{align}
\Delta({\cal W}_{j',m',m})=\Delta(e^{-\phi})+\Delta(\psi)-\frac{j(j-1)}{k}
+\frac{j'(j'+1)}{k'}=1 \,,
\end{align}
and since $\Delta(e^{-\phi})=1/2$, the conformal dimension of the
matter part is indeed $\Delta=\frac{1}{2}$, as required. 

Since $k'=N_5$, only $N_5-1$ of the $N$ $n$-cycle twist operators in
the orbifold theory have a dual worldsheet operator of the type
(\ref{wsW}); in the following we shall only consider those. 
It is generally believed that the remaining twist operators are dual
to worldsheet operators involving spectrally flowed $SL(2)$
representations \cite{Maldacena:2000hw,Argurio:2000tb}. 

\noindent Introducing the compact notation 
\begin{align}
 \psi^a=(\psi^+, \psi^0, \psi^-) \,,\qquad
 b_a = (-\textstyle\frac{1}{2},1,-\textstyle\frac{1}{2}) \, \qquad
 a=(+,0,-)\,.
\end{align}
the operator ${\cal W}_{j',m',m}$ can be written as 
\begin{align}  \label{W}
  {\cal W}_{j',m',m} = e^{-\phi} b_a \psi^a \Phi_{j,m-a}
  \Phi'_{j',m'} \ .
\end{align}
The analysis for the right-movers is identical, and the full local
fields are then \cite{Kutasov} (see also \cite{Giveon, GK}) 
\begin{align} \label{V1}
 V^{(1)}_{j',m',\bar m',m,\bar m} (z,\bar z)= {\cal W}_{j',m',m} (z)
\,  \bar {\cal W}_{j',\bar m',\bar m} (\bar z) \ .
\end{align}
The chiral primary $V^{(1)}_{j',m',\bar m',m,\bar m}$ is the `Fourier
transform' of the operator
\begin{align}
  V^{(1)}_{j',m',\bar m'}(z, \bar z; x,\bar x) =
  \sum_{m,\bar m} V^{(1)}_{j',m',\bar m',m,\bar m}(z,\bar z)\,
  x^{-m-j'-1} \bar x^{-\bar m-j'-1} \ ,
\end{align} 
which depends both on the worldsheet coordinates $(z, \bar z)$ as well
as on the representation labels $(x, \bar x)$. In addition to the 
worldsheet conformal dimension $(\Delta,\bar \Delta)=(1,1)$, the
operators $V^{(1)}_{j',m',\bar m'}(z, \bar z; x,\bar x)$ have also
spacetime scaling dimensions $(h,\bar h)=(j',j')$
\cite{Kutasov}. Since the variables $x$ and $\bar{x}$ are to be
identified with the complex coordinates of the 2d conformal field
theory on the boundary, these dimensions predict the scaling of the
two-point function of $V^{(1)}_{j',m',\bar m'}(z, \bar z; x,\bar x)$
as $|z_{12}|^{-4\Delta}$ and $|x_{12}|^{-4h}$. 
\medskip

For the analysis of the 3-point functions we also need the
corresponding operators with ghost number zero 
\begin{align}\label{V0}
V^{(0)}_{j',m',\bar m', m,\bar m} (z,\bar z) 
={\cal W}_{j',m',m}^0(z) \, 
\bar {\cal W}_{j',\bar m',\bar m}^0 (\bar z)
\ , 
\end{align}
that are obtained from $V^{(1)}_{j',m',\bar m',m,\bar m}$
by acting with the picture changing operator $\Gamma_{+1}$.
The only non-vanishing contribution comes from 
$e^\phi G_{-\frac{1}{2}}$ in $\Gamma_{+1}$, and hence one finds 
\begin{align}
{\cal W}^0_{j',m',m} = e^\phi G_{-\frac{1}{2}} {\cal W}_{j',m',m} \,,
\end{align}
where the global $\N=1$ superconformal generator
$G_{-\frac{1}{2}}=\frac{1}{2\pi i}\oint G(z)$ is given by the
supercurrent $G(z)$
\begin{align}
G(z)=\frac{2}{k} \left(\psi^a J_a -\frac{i}{3k} f_{abc} 
\psi^a \psi^b \psi^c \right) + \frac{2}{k} \left(\chi^a K_a -
\frac{i}{3k} f'_{abc} \chi^a \chi^b \chi^c \right) +
\lambda^i \partial Y_i \label{supercurrent}  \ .
\end{align} 
Using the $SL(2)$ structure constants $f_{abc}$, we can write the
$SL(2)$ part of $G(z)$ as
\begin{align}
G(z)\Bigr|_{SL(2)} = \frac{1}{k} \left(\psi^+J^- +
\psi^-J^+ - 2\psi^0 J^0 +\frac{2}{k} \psi^+\psi^0\psi^-
\right) \ .
\end{align} 
After some algebra, we find, more explicitly, 
\begin{align}
{\cal W}^0_{j',m',m} = b_a 
:J^a \Phi_{j,m-a}: \Phi'_{j',m'} +
\frac{2}{k} A_{ab} \psi^a\psi^b \Phi_{j,m-a-b} 
\Phi'_{j',m'} \equiv 
{\cal W}^{0,A}_{j',m',m} + {\cal W}^{0,B}_{j',m',m} 
 \ , \label{W0explicit}
\end{align}
where the matrix $A_{ab}= A^{(1)}_{ab}+ A^{(2)}_{ab}$ is given by
\begin{align}
A^{(1)}_{ab} = - d_a^{m-b} b_b \,,\qquad A^{(2)}_{ab} = 
\begin{pmatrix} 
0&b_{-}&0\\ 
0&0&b_{+}\\
-\textstyle\frac{1}{2}b_0 &0&0 
\end{pmatrix} \,, \qquad (a,b = +,0,-)
\end{align}
and
\begin{align}
b_a = (-\textstyle\frac{1}{2},1,-\textstyle\frac{1}{2})\,,\qquad
d_{a}^m = (-\textstyle\frac{1}{2}(j-1+m),m,\textstyle\frac{1}{2}(j-1-m))
\qquad(a=+,0,-) \ .
\end{align}
For later convenience we denote the two terms in
Eq.~(\ref{W0explicit}) by ${\cal W}^{0,A}_{j',m',m}$ and ${\cal
  W}^{0,B}_{j',m',m}$, respectively.  We note that in writing ${\cal
  W}^{0}_{j',m',m}$ we ignored terms coming the $SU(2)$ part of
$G(z)$, such as $\psi^a\chi^b \Phi_{j,m-a}\Phi'_{j',m'-b}$.  For
reasons discussed below, we do not need the precise form of these
terms. (The $U(1)$-part of $G_{-\frac{1}{2}}$ acts anyway trivial
since ${\cal W}$ is in the ground state with respect to the $U(1)$
excitations.)

\subsection{Correlators in the worldsheet theory}

It is suggested in \cite{Kutasov} that the operators
$V^{(1)}_{j',m',\bar m',m,\bar m}$ and $V^{(0)}_{j',m',\bar m',m,\bar
m}$ of the world-sheet theory correspond to the chiral primary
operators $V^{}_{j',m',\bar m'}$ of the boundary conformal field
theory. We now want to check this identification by comparing their
correlation functions.

\subsubsection{Worldsheet two-point function}

Let us begin with the two-point function of the chiral primary field 
$V^{(1)}_{j',m',\bar m',m,\bar m}(z, \bar z)$,
\begin{align} 
 G_2 = g_s^{-2} \langle V^{(1)}_{j',m',\bar m',m,\bar m}(z_1, \bar z_1)\, 
 V^{(1)}_{j',-m',-\bar m',-m,-\bar m}(z_2, \bar z_2) \rangle_{S^2} \ ,
\end{align} 
which will be evaluated on the sphere. Substituting the explicit
expression of $V^{(1)}_{j',m',\bar m',m,\bar m}$, Eq.~(\ref{V1}), into
$G_2$, we get 
\begin{align}
  G_2 &=g_s^{-2} b_{a_1} b_{\bar a_1} b_{a_2} b_{\bar a_2}
  \langle \psi^{a_1} \psi^{a_2} \rangle 
  \langle \bar \psi^{\bar a_1} \bar \psi^{\bar a_2}\rangle 
  {|z_{12}|^{-2}}\nonumber\\ 
  &\quad \times \langle \Phi_{j,m-a_1,\bar m-\bar a_1}  
  \Phi_{j,-m-a_2,-\bar m-\bar a_2}
  \rangle \langle \Phi'_{j',m',\bar m'}\Phi'_{j',-m',-\bar m'} \rangle
\ ,
\end{align}
where the summation over $a_1, \bar a_1, a_2, \bar a_2$ is
understood. The factor $|z_{12}|^{-2}$ comes from the 
propagator of the ghosts. Using the form of the
2-point function of the $SU(2)$ 
chiral primary fields $\Phi'_{j',m',\bar m'}$ given in
Eq.~(\ref{propphiprime}), and that of the fermions $\psi^a$, 
\begin{align} \label{proppsi}
 \langle \psi^{a_1}(z_1) \psi^{a_2}(z_2) \rangle = 
 \frac{k\, \eta^{a_1,a_2}}{2\, z_{12}}
 \qquad (a_1,a_2= 0,\pm 1) \ ,
\end{align} 
where $\eta^{+-}=2$ and $\eta^{00}=-1$, the two-point function $G_2$ 
simplifies to
\begin{align}
G_2 &=g_s^{-2} \frac{k^2}{4}
  \frac{b_{a_1} b_{\bar a_1} b_{a_2} b_{\bar a_2}} 
  {|z_{12}|^{4\Delta'_{j'}+4}}
  \eta^{a_1 a_2} \eta^{\bar a_1 \bar a_2}
  \langle \Phi_{j,m-a_1,\bar m-\bar a_1}  \Phi_{j,-m-a_2,-\bar m-\bar a_2}
  \rangle \ .
\end{align}
The only nonvanishing terms in $G_2$ are those for which $a_1=-a_2$
and $\bar a_1=-\bar a_2$. 

Next we employ the formula for the two-point function of the $SL(2)$
primaries in $m$-space, 
\begin{align}
\langle\Phi_{j,m,\bar m} \Phi_{j,-m,-\bar m} \rangle
&= \frac{\pi B(j) \delta(0)}{|z_{12}|^{4\Delta_j}}
\frac{\Gamma(1-2j)\Gamma(j-m)\Gamma(j+\bar m)}
{\Gamma(2j)\Gamma(1-j-m)\Gamma(1+\bar m-j)} \ , \label{2ptdelta}
\end{align}
with $B(j)$ as defined in Eq.~(\ref{Bj}).  Then we sum over $a_1$ and
$\bar a_1$ (nine terms) and obtain
\begin{align}
  G_2&= g_s^{-2} \frac{k^2}{16} \frac{\pi B(j) \delta(0)}
{|z_{12}|^{4(\Delta_j+\Delta'_{j'}+{1})}}
\frac{\Gamma(1-2j')\Gamma(j'-m)\Gamma(j'+\bar m)}
{\Gamma(2j')\Gamma(1-j'-m)\Gamma(1+\bar m-j')} \ .   
\end{align}
It is interesting to observe that in comparison with the two-point
function (\ref{2ptdelta}), the arguments in the gamma functions in
$G_2$ are shifted from $j$ to $j'=j-1$. This shows that in the
definition of the chiral primary the action of $\psi^a$ on the $SL(2)$
operators $\Phi_{j,m,\bar m}$ shifts the $x$-dependence of the
two-point function from $|x_{12}|^{-4j}$ to $|x_{12}|^{-4(j-1)}$.
Transforming $G_2$ back to coordinate space, we obtain
\begin{align}
G_2 &= g_s^{-2} \frac{k^2}{16}  \frac{B(j) \delta(0)}
{|x_{12}|^{4(j-1)}}    
\frac{1}{|z_{12}|^{4}} \,\label{result2pt} 
\end{align}
which has the expected scaling behavior, $|z_{12}|^{-4\Delta}$ and
$|x_{12}|^{-4h}$ ($\Delta=1$, $h=j'$). 

In order to obtain the two-point function of the corresponding
$n$-cycle twist operators of the boundary conformal field theory from
this, we now follow the prescription given in~\cite{Maldacena}. We
find 
\begin{align}\label{relation2pt}
&\langle V^{}_{j',m',\bar m'}(x_1, \bar x_1) \, 
V^{}_{j',-m',-\bar m'}(x_2,\bar x_2)
  \rangle_{\rm bCFT} \nonumber\\
  &\quad= \frac{1}{V_{\rm conf}} g_s^{-2} \,
\langle V^{(1)}_{j',m',\bar m'}(z_1=\bar z_1=1;x_1, \bar x_1) \, 
V^{(1)}_{j',-m',-\bar m'}(z_2=\bar z_2=0; x_2, \bar x_2) 
\rangle_{S^2} 
 \nonumber \\
 &\quad= g_s^{-2}\, \frac{k^2}{16}\,   \frac{B(j)}{|x_{12}|^{4(j-1)}} \, 
\frac{\delta(0)}{V_{\rm conf}} \nonumber \\
&\quad = 
(2j-1)\, g_s^{-2}\, \frac{k^2}{16} \, \frac{B(j)}{|x_{12}|^{4j'}} \ .
\end{align}
Here $V_{\rm conf}=\int d^2z |z|^{-2}$ is the volume of the conformal
group on the sphere (the M\"obius group); this factor cancels the
divergence coming from the delta function $\delta(0)$ in
Eq.~(\ref{result2pt}) up to a $j$-dependent factor, which is $2j-1$
\cite{Maldacena}.  

Now we can compare this result to the boundary two-point function
(\ref{2ptbCFT}) found in the symmetric orbifold theory. In particular,
this now fixes the relative normalisation between the worldsheet
vertex operators $V^{(1)}_{j',m',\bar m'} (z,\bar z; x, \bar x)$ and
the conformal field theory operators $V^{}_{j',m',\bar m'} (x,\bar x)$
\begin{align}\label{resca}
V^{(1)}_{j',m',\bar m'} (z,\bar z; x, \bar x) 
\Longleftrightarrow a(j) \, V^{}_{j',m',\bar m'} (x,\bar x) \ , 
\end{align}
where 
\begin{align}
a(j) = \frac{k}{4g_s} \sqrt{(2j-1) B(j)}   \ .
\end{align} 

\subsubsection{Worldsheet three-point function } \label{sec3pt}

We now consider the three-point function 
\begin{align} \label{3ptV1V1V0}
G_3 = g_s^{-2} \left\langle
     V^{(1)}_{j'_1,m'_1,\bar m'_1,m_1,\bar m_1} (z_1,\bar z_1) \,
     V^{(1)}_{j'_2,m'_2,\bar m'_2,m_2,\bar m_2} (z_2,\bar z_2) \,
     V^{(0)}_{j'_3,m'_3,\bar m'_3,m_3,\bar m_3} (z_3,\bar z_3) 
      \right\rangle_{S^2} \ ,
\end{align}
where $V^{(0)}_{j',m',\bar m',m,\bar m}$ is the descendant of the
chiral primary $V^{(1)}_{j',m',\bar m',m,\bar m}$. Since $V^{(1)}$ has
ghost number $-1$, while $V^{(0)}$ has ghost number zero, the total
ghost number of $G_3$ is then $-2$, as required on the sphere. 
The explicit calculation can now be divided into four
parts 
\begin{align} \label{G3terms}
G_3 = G_3^{AA} + G_3^{AB}+ G_3^{BA} + G_3^{BB} \ ,
\end{align}
where
\begin{align}
G_3^{AA} &=g_s^{-2} \left\langle
      V^{(1)}_{j'_1,m'_1,\bar m'_1,m_1,\bar m_1} (z_1,\bar z_1) \,
      V^{(1)}_{j'_2,m'_2,\bar m'_2,m_2,\bar m_2} (z_2,\bar z_2) \,
      {\cal W}^{0,A}_{j'_3,m'_3,m_3} (z_3) \,
      \bar {\cal W}^{0,A}_{j'_3,\bar m'_3,\bar m_3} 
     (\bar z_3) 
     \right\rangle  \nonumber\\
G_3^{AB} &=g_s^{-2} \left\langle
      V^{(1)}_{j'_1,m'_1,\bar m'_1,m_1,\bar m_1} (z_1,\bar z_1) \,
      V^{(1)}_{j'_2,m'_2,\bar m'_2,m_2,\bar m_2} (z_2,\bar z_2)\,
      {\cal W}^{0,A}_{j'_3,m'_3,m_3} (z_3) \, 
      \bar {\cal W}^{0,B}_{j'_3,\bar m'_3,\bar m_3} (\bar z_3) 
     \right\rangle  \nonumber\\
G_3^{BA} &=g_s^{-2} \left\langle
      V^{(1)}_{j'_1,m'_1,\bar m'_1,m_1,\bar m_1} (z_1,\bar z_1) \,
      V^{(1)}_{j'_2,m'_2,\bar m'_2,m_2,\bar m_2} (z_2,\bar z_2) \,
      {\cal W}^{0,B}_{j'_3,m'_3,m_3} (z_3) \,
      \bar {\cal W}^{0,A}_{j'_3,\bar m'_3,\bar m_3} 
     (\bar z_3) 
     \right\rangle \nonumber\\
G_3^{BB} &=g_s^{-2} \left\langle
      V^{(1)}_{j'_1,m'_1,\bar m'_1,m_1,\bar m_1} (z_1,\bar z_1) \,
      V^{(1)}_{j'_2,m'_2,\bar m'_2,m_2,\bar m_2} (z_2,\bar z_2) \,
      {\cal W}^{0,B}_{j'_3,m'_3,m_3} (z_3) \,
      \bar {\cal W}^{0,B}_{j'_3,\bar m'_3,\bar m_3} 
     (\bar z_3)  
     \right\rangle \ , \nonumber 
\end{align}
and ${\cal W}^{0,A}$ and ${\cal W}^{0,B}$ are defined as in 
Eq.~(\ref{W0explicit}).

We begin with the correlator $G_3^{BB}$. Substituting the explicit
expressions for the vertex operators, we get 
\begin{align} \label{G3BB}
G^{BB}_3 &=g_s^{-2} \frac{4}{k^2}
b_{a_1} b_{a_2} A_{a_3b_3} b_{\bar a_1} b_{\bar a_2}A_{\bar a_3 \bar b_3} 
\langle \psi^{a_1}\psi^{a_2}\psi^{a_3} \psi^{b_3} \rangle
\langle \bar \psi^{\bar a_1}\bar \psi^{\bar a_2}
        \bar \psi^{\bar a_3}\bar \psi^{\bar b_3} \rangle\nonumber \\
&\quad\times \langle \Phi_{j_1,m_1-a_1, \bar m_1-\bar a_1}
\Phi_{j_2,m_2-a_2,\bar m_2-\bar a_2}
\Phi_{j_3,m_3-a_3-b_3,\bar m_3-\bar a_3-\bar b_3} \rangle \nonumber\\
&\quad\times\langle  \Phi'_{j'_1,m'_1, \bar m'_1}\Phi'_{j'_2,m'_2,\bar m'_2}
\Phi'_{j'_3, m'_3,\bar  m'_3} \rangle  {|z_{12}|^{-2}} \,,
\end{align}
where summation over $a_i,b_3, \bar a_i,\bar b_3$ $(i=1,2,3)$ is
understood.  Here we have already omitted terms in the operator
$V^{(0)}_{j',m',\bar m',m,\bar m}$ which involve the spinor $\chi^a$
({\it cf.}\ the discussion at the end of section~3.1).  Since the
$SU(2)$ spinor $\chi^a$ cannot be contracted with the $SL(2)$ spinor
$\psi^a$, such terms do not contribute to the three-point function.
The factor $|z_{12}|^{-2}$ is as before the contribution from the
ghosts. 

Let us consider in detail the three-point functions of the primary
fields $\psi^a$, $\Phi_{j,m,\bar m}$ and $\Phi'_{j',m',\bar m'}$.  The
three-point function involving the fermions $\psi^a$ in $G^{BB}_3$ is
evaluated using Wick contraction and the fermion propagator
(\ref{proppsi})
\begin{align} \label{3ptpsi}
\langle \psi^{a_1}(z_1) \psi^{a_2} (z_2) \psi^{a_3}(z_3) 
\psi^{b_3}(z_3) \rangle
=k^2 \frac{\eta^{a_1[b_3} \eta^{a_3]a_2}}
{z_{13}z_{23}} \ .
\end{align}
This is non-vanishing only if 
\begin{align} \label{constraint1}
(a_1 = -b_3\ , \; a_2=-a_3 \ , \; a_3 \neq b_3)\qquad
\hbox{or} \qquad 
(a_1 = -a_3\ , \; a_2=-b_3 \ , \; a_3 \neq b_3) \ .
\end{align}
These constraints imply the relation $a_1+a_2+a_3+b_3=0$.

The three-point functions of the $SL(2)$ and $SU(2)$ primary fields 
$\Phi_{j,m,\bar m}$ and $\Phi'_{j',m',\bar m'}$ are given by
\begin{align}
&\langle \Phi_{j_1,m_1,\bar m_1}(z_1,\bar z_1) \, 
\Phi_{j_2,m_2,\bar m_2}(z_2,\bar z_2) \, 
\Phi_{j_3,m_3,\bar m_3} (z_3,\bar z_3)\rangle \nonumber\\ 
&\qquad \qquad \qquad= \delta^2(m_1+m_2+m_3) \,
W(j_i;m_i) \, C_{j_1, j_2, j_3} \, 
\prod_{i<j} \frac{1} {|z_{ij}|^{2\Delta_{ij}}}\label{3ptphi}
\\
\langle &  \Phi'_{j'_1,m'_1, \bar m'_1}(z_1,\bar z_1) \,
\Phi'_{j'_2,m'_2,\bar m'_2}(z_2,\bar z_2) \,
\Phi'_{j'_3, m'_3,\bar  m'_3} (z_3,\bar z_3)\rangle  \nonumber\\ 
&\qquad \qquad \qquad =  \delta^2(m'_1+m'_2+m'_3) \,
 \hat W(j'_i;m'_i) \, C'_{j'_1, j'_2, j'_3} \,
\prod_{i<j} \frac{1} {|z_{ij}|^{2\Delta'_{ij}}} \label{3ptphi'} \ ,
\end{align}
where $C_{j_1,j_2,j_3}$ and $C'_{j'_1,j'_2,j'_3}$ are the $SL(2)$ and
$SU(2)$ structure constants, respectively, that are explicitly given
in Eqns.~(\ref{CSL2}) and (\ref{CSU2}) of appendix~\ref{AppA}. We have
used the conventions $x_{12}=x_1-x_2$, $z_{12}=z_1-z_2$,
$\Delta_{12}=\Delta_{j_1}+\Delta_{j_2} - \Delta_{j_3}$,
$j_{12}=j_1+j_2-j_3$, {\it etc}. The $\delta^{2}$ symbol means
that the three-point function (\ref{3ptphi}) is only nonvanishing for
$m_1+m_2+m_3=\bar m_1+\bar m_2+\bar m_3=0$, and similarly for 
(\ref{3ptphi'}).

The function $W(j_i,m_i)$ is the `Fourier transform' of the factor
$\prod_{i<j} {|x_{ij}|^{-2j_{ij}}}$, {\it i.e.}\ it reflects the
dependence on the representation label $x$. This function is given by
the integral
\begin{align} \label{Wj}
W(j_i;m_i) &= \int d^2 x_2\, d^2 x_3\, 
x_2^{j_2+m_2-1} \bar x_2^{j_2+\bar m_2-1} 
|1-x_2|^{-2j_{12}} \\ 
&\quad\quad \times x_3^{j_3+m_3-1} \bar x_3^{j_3+\bar m_3-1} 
|1-x_3|^{-2j_{13}} |x_2-x_3|^{-2j_{23}} \ . \nonumber
\end{align}
An explicit expression for this integral has been found by Satoh
\cite{Satoh},\footnote{In Satoh's analysis this integral appears in a
slightly different context as he uses different conventions for the
$SL(2)$ representations.} 
\begin{align} 
&W(j_i;m_i) = (-)^{w} \frac{\pi^2 \gamma(-\tilde N) \gamma(2j'_3+1)}
{\gamma(1+j'_{31})\gamma(1+j'_{32})} 
\frac{\Gamma(1+j'_2-m_2)\Gamma(1+j'_2-\bar m_2)}
{\Gamma(1+j'_2-m_2-n_3)\Gamma(1+j'_2-\bar m_2-\bar n_3)} \nonumber\\
&\quad\times \prod_{a=1,2} 
\frac{\Gamma(1+j'_a+m_a)}{\Gamma(-j'_a-\bar m_a)}
F \begin{bmatrix} -n_3, -j'_{31}, 1+j'_{12} \\ 
-2j'_3, 1+j'_2-m_2-n_3 \end{bmatrix}
F \begin{bmatrix} -\bar n_3, -j'_{31}, 1+j'_{12} \\ 
-2j'_3, 1+j'_2-\bar m_2-\bar n_3 
\end{bmatrix} \!\!\!
\end{align}
with $w=m_2-\bar m_2+\bar n_3$ and $j'_i=j_i-1$.  Furthermore,
$m_3=-j'_3+n_3$ with $n_3 \geq 0$, and $\bar m_3=-j'_3+\bar n_3$ with 
$\bar n_3 \geq 0$ (see appendix~A). Finally, $F[a,b,c;e,f]$ is the
hypergeometric function ${}_3F_2(a,b,c;e,f;1)$ and 
$\tilde N=j'_1+j'_2+j'_3+1$. 

We want to compare this 3-point function with the 3-point function of
the boundary conformal field theory (\ref{mprimes}). There we
restricted ourselves to the case (\ref{mprimes}) with 
$d=j'_{12} \geq 0$. For these special values also the function 
$\hat W(j'_i;m'_i)$ simplifies, and we find
\begin{align}
\hat W(j'_i; m'_i)= 
\frac{\Gamma(j'_{13}+1)\Gamma(j'_{12}+1)}{\Gamma(2j'_1+1)} 
\ .
\end{align}
This is shown in appendix~\ref{AppA}, using the $SU(2)$ OPE
coefficients \cite{Dotsenko} for the case described by
(\ref{mprimes}). 
Substituting Eqs.~(\ref{3ptpsi}), (\ref{3ptphi}) and (\ref{3ptphi'})
in $G^{BB}_3$, we obtain
\begin{align}
G^{BB}_3 &= g_s^{-2} k^2\, 
b_{a_1} b_{a_2} A_{a_3b_3} b_{\bar a_1} b_{\bar a_2}A_{\bar a_3 \bar b_3}
 \,\eta^{a_1[b_3} \eta^{a_3]a_2} \eta^{\bar a_1[\bar b_3} \eta^{\bar a_3]
 \bar a_2} \,C_{j_1, j_2, j_3}C'_{j'_1, j'_2, j'_3}
\delta^2({\textstyle \sum_i m_i})\nonumber\\
&\times 
W(j_i;m_1-a_1,m_2-a_2,m_3-a_3-b_3) \hat W(j'_i;m'_i) 
\prod_{i<j}\frac{1}{|z_{ij}|^{2(\Delta_{ij}+\Delta'_{ij}+1)}} \,.
\end{align}

We will now perform the sum over the eight indices $a_i,b_3, \bar
a_i,\bar b_3$ ($i=1,2,3$). We expect that after taking the sum, the
function $W(j_i; \ldots)$ in $G^{BB}_3$ is shifted to $W(j_i-1; m_i)$,
which is the `Fourier transform' of $\prod_{i<j}
{|x_{ij}|^{-2j_{ij}+2}} = \prod_{i<j} {|x_{ij}|^{-2j'_{ij}}}$.
$G^{BB}_3$ would then have the same dependence on the coordinates
$x_i$ ($i=1,2,3$) as the boundary three-point
function~(\ref{3ptbCFT}). In order to check this, we consider the
special case where the $SL(2)$ quantum numbers are\footnote{Because of
the $SL(2)$ covariance, this result should then hold for arbitrary
values of  $m_i$, $\bar m_i$ ($i=1,2,3$).}
\begin{align} \label{condm3}
  m_1&=\bar m_1 =j'_1-d \ , \nonumber\\
  m_2&=\bar m_2 =j'_2-1 \ , \\
  m_3&=\bar m_3 =-j'_3+1=-(j'_1+j'_2-d)+1 \ . \nonumber
\end{align}
The condition that $n_3\geq 0$ then translates into 
\begin{align} \label{constraint2}
n_3=1-a_3-b_3 \geq 0 \ .
\end{align}
The constraints (\ref{constraint1}) and (\ref{constraint2}) (and
similar constraints for the bared indices) imply that there are $144$
nonvanishing terms in $G^{BB}_3$ which we sum up using computer
algebra. We then obtain
\begin{align}
G^{BB}_3 = g_s^{-2} g^{BB}(j'_i) \,C_{j_1, j_2, j_3}\,
C'_{j'_1, j'_2, j'_3} \, W(j_i-1;m_i) \, \hat W(j'_i;m'_i) \, 
\prod_{i<j}\frac{1}{|z_{ij}|^{2(\Delta_{ij}+\Delta'_{ij}+1)}} \ ,
\end{align}
where the function $g^{BB}(j'_i)$ turns out to be
\begin{align} \label{fji}
g^{BB}(j'_i) &= \frac{k^2}{16} {{(2j'_3)}^2} \ .
\end{align}
In particular, this expression shows the desired shift in the $j_i$
dependence of $W$. 

We proceed similarly for the remaining terms $G^{AA}_3$, $G^{AB}_3$
and $G^{BA}_3$. As shown in appendix~\ref{AppB}, these terms have the
same structure as $G^{BB}_3$, but with $g^{BB}(j'_i)$ replaced by
\begin{align} 
g^{AA}(j'_i) &= \frac{k^2}{16}\,  {(1 + j_1' + j_2' - j_3')^2}  \ ,\\
g^{AB}(j'_i)&=g^{BA}(j'_i) 
= \frac{k^2}{16} \, (1 + j_1' + j_2' - j_3') 2 j_3' \ .
\end{align}
After adding up the four terms, we obtain
\begin{align}
G_3 = g_s^{-2} \, g(j'_i) \, C_{j_1, j_2, j_3}\, C'_{j'_1, j'_2, j'_3} \, 
W(j_i-1;m_i) \, \hat W(j'_i;m'_i)  \,
\prod_{i<j}\frac{1}{|z_{ij}|^{2(\Delta_{ij}+\Delta'_{ij}+1)}} \ ,
\end{align}
with
\begin{align} 
g(j'_i) &=  g^{AA}(j'_i) + g^{AB}(j'_i) + g^{BA}(j'_i) + g^{BB}(j'_i)
=\frac{k^2}{16} \, {(1+j'_1+j'_2+j'_3)^2} \ .
\end{align}

The product of the $SL(2)$ and $SU(2)$ structure constants
$C_{j_1,j_2,j_3}$ and $C'_{j'_1,j'_2,j'_3}$ in $G_3$ can now be
simplified using the identity
\begin{align} \label{identityCC}
C_{j_1,j_2,j_3}C'_{j'_1,j'_2,j'_3}= 
\sqrt{ B(j_1)B(j_2)B(j_3)}\ ,
\end{align}
which is shown in appendix~\ref{appendixproduct}. This finally allows
us to write $G_3$ as
\begin{align}
G_3 = g_s^{-2} \, \frac{k^2}{16} \, {(j'_1+j'_2+j'_3+1)^2} \,
W(j_i-1;m_i) \, \hat W(j'_i;m'_i) \,
\prod_i B(j_i)^{1/2}
\prod_{i<j}\frac{1}{|z_{ij}|^{2}} \ .
\end{align}

Before comparing this to the dual boundary conformal field theory 
answer let us pause to comment on the identity (\ref{identityCC})
which is somewhat striking: it states that, for canonically
normalised fields, the operator algebra coefficients of 
$SL(2)_k$ are inverse to those of $SU(2)_{k'}$ (for $k=k'$)! The
technical reason for this is that the functions $G(j)$ appearing in
$C_{j_1,j_2,j_3}$ and the functions $P(j')$ occurring in
$C'_{j'_1,j'_2,j'_3}$ behave inversely to one another. More precisely,
from the definition of the functions $G(j)$ and $P(j')$, we obtain the
simple relation
\begin{align}
 G(-j)=\frac{G(-1)}{P(j')} \, ,
\end{align}
where $G(-1)$ is a regular function, see
appendix~\ref{appendixproduct} for details.  In the product $C \cdot
C'$ the poles of the structure constants $C_{j_1,j_2,j_3}$ therefore
cancel precisely against the zeros of $C'_{j'_1,j'_2,j'_3}$.
\smallskip

Returning to the question of the AdS/CFT correspondence, we need to
transform $G_3$ back into $x$-space in order to compare it with the
boundary three-point function (\ref{3ptbCFT}). This is now trivial
since we can use the Satoh formula again, except that now $j_i$ has
been shifted to $j_i'=j_i-1$, as already discussed above. In
particular, this therefore reproduces the same $x$-dependence as in 
(\ref{3ptbCFT}). It thus remains to check that also the overall factor
of the three-point functions agree. From the analysis of the two-point
function we have deduced how the fields have to be rescaled, see
Eq.~(\ref{resca}). Taking this into account, we then obtain from $G_3$
the rescaled function
\begin{align}
  \widehat{G}_3 &=
  \frac{4 g_s}{k}\, (j'_1+j'_2+j'_3+1)^2\,
  \frac{\Gamma(j'_{13}+1)\Gamma(j'_{12}+1)}{\Gamma(2j'_1+1)}\,
  \prod_i (2j'_i+1)^{-1/2}\,
  \prod_{i<j} \frac{1}{|x_{ij}|^{2j'_{ij}}}\,,
\end{align}
where we have also performed the integral over the world-sheet
coordinates, which in this case just cancels the volume of the
M\"obius group. Since $k=N_5$, $\widehat{G}_3$ scales as
$g_s/k \sim 1/\sqrt{N_1N_5} = 1/\sqrt{N}$. This large $N$ scaling
behavior agrees then precisely with that of the boundary three-point
function (\ref{3ptbCFT}), see (\ref{Nlim}). This is a non-trivial
consistency check since the functional dependence on the $j'_i$ is
quite complicated!


%
\section{Conclusions}
%

In this paper we have compared correlation functions of chiral
primary operators of the $AdS_3\times S^3 \times T^4$ WZW model with
the corresponding amplitudes in the boundary conformal field theory
that can be defined as a symmetric orbifold. The comparison of the
2-point functions determines the relative normalisation of the chiral
primary fields in the two descriptions. It is then a non-trivial
consistency check to compare the coefficients of the 3-point
functions. In the large $N$ limit in which the sphere correlators
(that we have calculated) dominate the string perturbation series, we
have found beautiful agreement. We should note that the chiral
primaries we have considered lie in short multiplets and are therefore
protected by non-renormalisation theorems. It therefore makes sense to
compare these correlation functions. 

It would be interesting, though technically demanding, to repeat this
analysis for the 4-point functions. This is the first example where
in the world-sheet theory a non-trivial integral over the cross-ratio
will have to be performed. It would be interesting to understand in 
detail how this will manifest itself in the dual boundary conformal
field theory. For the case of the usual $AdS_5\times S^5$ case,
Gopakumar has suggested \cite{Gopakumar} that this integral becomes
the Schwinger parametrisation of the propagator. This idea has
recently been tested a little bit further \cite{Aharony,David2}.

\bigskip

\noindent {\it Note added:} After completion of this paper the paper 
\cite{after} appeared which contains some overlapping results.

%
\section*{Acknowledgments}
%

We would like to thank Marco Baumgartl, Finn Larsen, Andreas Ludwig
and J\"org Teschner for helpful discussions and correspondences. 
M.R.G.\ also thanks Amit Giveon for putting him straight about many
aspects of the AdS$_3$/CFT$_2$ correspondence. This work is partially
supported by the Swiss National Science Foundation and the Marie Curie
network `Constituents, Fundamental Forces and Symmetries of the
Universe' (MRTN-CT-2004-005104). 

\bigskip
\bigskip

\appendix

%
\begin{Large} \noindent \bf Appendix \end{Large}
%

\section{Correlators in $SL(2)_k$ and $SU(2)_{k'}$ WZW models} 
\label{AppA}

\subsection{Two- and three-point functions in the $SL(2)_k$ WZW model}

The chiral primaries of the $SL(2)$ WZW model are denoted 
by\footnote{In this appendix we only deal with the bosonic currents;
$k$ and $k'$ therefore refer to the bosonic levels.}
\begin{align}
\Phi_{j,m,\bar m}(z,\bar z)=\Phi_{j,m}(z) \, 
\bar \Phi_{j,\bar m}(\bar z) 
\qquad \hbox{with} \qquad 
\Delta_j =\bar\Delta_j =-\frac{j(j-1)}{k-2} \ ,
\end{align}
where $k$ is the level of the affine Lie algebra. In the current
context only half-integer $j$ will be relevant (because of
(\ref{jj'})). In this case, the OPEs (\ref{currents}) imply that the
values of $m$ and $\bar{m}$ run between $m=-(j-1),\ldots,(j-1)$.    

\noindent The vertex operators $\Phi_{j,m,\bar m}(z,\bar z)$ are the 
`Fourier transforms' of the operators $\Phi_j(z, \bar z; x,\bar x)$
\begin{align}
\Phi_{j,m,\bar m} (z, \bar z)= 
\int d^2x\, x^{j+m-1} \bar x^{j+\bar m-1} \Phi_j(z, \bar z; x,\bar x)
\ .
\end{align}
The inverse transformation is 
\begin{align} \label{invtrafo}
\Phi_{j}(z, \bar z; x,\bar x) = 
\frac{1}{V_{\rm conf}}
\sum_{m,\bar m} \Phi_{j,m,\bar m}(z,\bar z)\, x^{-m-j} 
\bar x^{-\bar m-j} \ ,
\end{align}
where $V_{\rm conf}=\int d^2x |x|^{-2}$.

The two- and three-point functions of $\Phi_j (z,\bar z; x,\bar x)$
were computed in \cite{Teschner1997, Teschner1999, FZZ}. The two-point
function is given by\footnote{The $SL(2)$ amplitudes were only
determined up to an overall normalisation in~\cite{Teschner1999}. 
In the following we shall use a different overall normalisation
from \cite{Teschner1999}, which seems to be more natural in the
current context; this will become apparent in appendix~C. 
We thank J\"org Teschner for explaining this
to us.}
\begin{align} \label{twopointsl2}
&\langle \Phi_{j_1}(z_1,\bar z_1; x_1,\bar x_1)
\Phi_{j_2}(z_2,\bar z_2;x_2,\bar x_2) \rangle  \nonumber\\
&\qquad = \frac{1}{|z_{12}|^{4\Delta_{j_1}}} 
\left[ \frac{1}{(2\pi)^2} \,\delta(x_{12}) \, \delta(\bar{x}_{12})\, 
\delta(j_1+j_2-1) + 
\frac{B(j_1)}{|x_{12}|^{4j_1}} \delta(j_1-j_2) \right] \ ,
\end{align}
with coefficient
\begin{align} \label{Bj}
B(j)= \frac{1}{(2\pi)^2} 
\frac{k-2}{\pi} \frac{\nu^{1-2j}}{\gamma(\frac{2j-1}{k-2})}
\qquad \hbox{and} \qquad 
\gamma(x)=\frac{\Gamma(x)}{\Gamma(1-x)} \ , \qquad \nu=\pi 
\frac{\Gamma(1-\frac{1}{k-2})}{\Gamma(1+\frac{1}{k-2})} \ .
\end{align}

\noindent The three-point function is 
\begin{align}
&\langle \Phi_{j_1}(z_1,\bar z_1;x_1,\bar x_1) \, 
         \Phi_{j_2}(z_2,\bar z_2;x_2,\bar x_2) \, 
         \Phi_{j_3}(z_3,\bar z_3;x_3,\bar x_3) \rangle
= C_{j_1j_2j_3}\, \prod_{i<j} 
\frac{1}{|x_{ij}|^{2j_{ij}}|z_{ij}|^{2\Delta_{ij}}}
\ ,
\end{align}
with $\Delta_{12}=\Delta_{j_1}+\Delta_{j_2}-\Delta_{j_3}$,
$j_{12}=j_1+j_2-j_3$, {\it etc.}\ and coefficients
\begin{align} \label{CSL2}
C_{j_1,j_2,j_3} = \frac{1}{(2\pi)^2}\,  \frac{k-2}{2\pi^3}\, 
 \frac{G(1-j_1-j_2-j_3) G(-j_{12}) G(-j_{23})G(-j_{31})}
 { \nu^{j_1+j_2+j_3-2} G(-1) G(1-2j_1)G(1-2j_2)G(1-2j_3)}\ ,
\end{align}
where
\begin{align} 
G(j)=(k-2)^{\frac{j(k-1-j)}{2(k-2)}} \, \Gamma_2(-j|1,k-2)  \,
\Gamma_2 (k-1+j|1, k-2) \ ,
\end{align}
and $\Gamma_2 (x|1, \omega)$ is the Barnes double Gamma function.
$G(j)$ has poles at $j=n+m(k-2)$ and $j=-n-1-(m+1)(k-2)$ with
$n,m=0,1,...$. A discussion of these poles can be found in
\cite{Maldacena}. In $C_{j_1,j_2,j_3}$ the poles $j_1+j_2+j_3=n+k$,
$n=0,1,...$ are excluded by the condition
\begin{align} \label{condj}
j_1+j_2+j_3 \leq k-1 \,.
\end{align}
The function $G(j)$ satisfies the recursion relation
\begin{align}
G(j+1)=\gamma(-\textstyle\frac{j+1}{k-2}) \, 
G(j) \ . \label{Gplus1}
\end{align}

\noindent In $m$-space the two- and three-point functions are given
by
\begin{align} \label{2ptmspace}
\langle\Phi_{j,m,\bar m} (z_1,\bar z_1) 
\Phi_{j,-m,-\bar m} (z_2,\bar z_2)  \rangle
&= \frac{\pi B(j)\delta(0)}{|z_{12}|^{4\Delta_j}}  
\frac{\Gamma(1-2j)\Gamma(j-m)\Gamma(j+\bar m)}
{\Gamma(2j)\Gamma(1-j-m)\Gamma(1+\bar m-j)} \,
\end{align}
and
\begin{align} \label{3ptmspace}
&\langle \Phi_{j_1,m_1,\bar m_1}(z_1,\bar z_1)
\Phi_{j_2,m_2,\bar m_2}(z_2,\bar z_2)
\Phi_{j_3,m_3,\bar m_3} (z_3, \bar z_3)\rangle \nonumber\\ 
&\qquad= \delta^2(m_1+m_2+m_3) 
W(j_a;m_a) C_{j_1, j_2, j_3} \prod_{i<j} \frac{1}
{|z_{ij}|^{2\Delta_{ij}}} \ ,
\end{align} 
with coefficients $B(j)$ and $C_{j_1,j_2,j_3}$ as above.  

\noindent An explicit expression for $W(j_i, m_i)$ can be found in
\cite{Satoh}.  Defining $j'_i=j_i-1$ ($i=1,2,3$), the integral
(\ref{Wj}) is identical to Eq.~(2.8) in \cite{Satoh} (upon cyclic
permutation of the indices) which can written as
\begin{align} \label{Wjama}
&W(j_i;m_i) = (-)^{m_2-\bar m_2+\bar n_3} \frac{\pi^2 
\gamma(-\tilde  N) \gamma(2j'_3+1)}
{\gamma(1+j'_{31})\gamma(1+j'_{32})} 
\frac{\Gamma(1+j'_2-m_2)\Gamma(1+j'_2-\bar m_2)}
{\Gamma(1+j'_2-m_2-n_3)\Gamma(1+j'_2-\bar m_2-\bar n_3)} \nonumber\\
&\quad\times \prod_{a=1,2} 
\frac{\Gamma(1+j'_a+m_a)}{\Gamma(-j'_a-\bar m_a)}
F \begin{bmatrix} -n_3, -j'_{31}, 1+j'_{12} \\ 
-2j'_3, 1+j'_2-m_2-n_3 
\end{bmatrix}
F \begin{bmatrix} -\bar n_3, -j'_{31}, 1+j'_{12} \\ 
-2j'_3, 1+j'_2-\bar m_2-\bar n_3 
\end{bmatrix}
\end{align}
where $F\begin{bmatrix} a,b,c \\ e,f \end{bmatrix}$ is the
hypergeometric function ${}_3F_2(a,b,c;e,f;1)$.  Here $\tilde
N=j'_1+j'_2+j'_3+1$, $m_3=-j'_3+n_3$ and $\bar m_3=-j'_3+\bar n_3$
($n_3,\bar n_3 =0,1,...$); $m_1, m_2$ and $\bar m_1, \bar m_2$
are arbitrary.

\subsection{Two- and three-point functions in the 
$SU(2)_{k'}$ WZW model}

The chiral primaries of the $SU(2)_{k'}$ WZW model are denoted by
\begin{align}
\Phi'_{j',m',\bar m'}(z,\bar z)=\Phi'_{j',m'}(z) \, 
\bar \Phi'_{j',\bar m'}(\bar z) \ ,
\end{align}
and have conformal dimension
\begin{align} 
\Delta'_{j'} = \bar{\Delta}'_{j'}=\frac{j'(j'+1)}{k'+2} \ ,
\qquad 0 \leq j' \leq \frac{k'}{2} \ ,
\end{align}
where $j'$ is the $SU(2)$ representation label and $k'$ the level of
the affine Lie algebra.

As for the case of $SL(2)$ it is convenient to introduce instead of
the $m'$ variables continuous $y$ variables (see for example
\cite{DAK}). In these conventions the two- and three-point functions
of $\Phi'_{j'}(z,\bar z; y, \bar y)$ are then 
\cite{Zamolodchikov,Dotsenko,DAK}
\begin{align} \label{propphiprime}
\langle \Phi'_{j'_1}(z_1,\bar z_1; y_1, \bar y_1) 
\Phi'_{j'_2}(z_2,\bar z_2; y_2, \bar y_2)
\rangle =\delta_{j'_1,j'_2}\, 
\frac{|y_{12}|^{2j'_1}}{|z_{12}|^{4\Delta'_{j'_1}}} \ ,
\end{align}
and 
\begin{align} \label{3ptphiprime}
\langle &\Phi'_{j'_1}(z_1,\bar z_1; y_1, \bar y_1)
\Phi'_{j'_2}(z_2, \bar z_2; y_2, \bar y_2) 
\Phi'_{j'_3}(z_3, \bar z_3; y_3 \bar y_3) \rangle 
= C'_{j'_1, j'_2, j'_3} \, \prod_{i<j} \frac{|y_{ij}|^{2j'_{ij}}}
{|z_{ij}|^{2\Delta'_{ij}}} \ , 
\end{align}
with $\Delta'_{12}=\Delta'_{j'_1}+\Delta'_{j'_2}-\Delta'_{j'_3}$, 
{\it etc.} The relevant coefficients are
\begin{align} \label{CSU2}
C'_{j'_1,j'_2,j'_3} = 
\sqrt{\frac{\gamma({\textstyle\frac{1}{k'+2}})}{
\gamma(\frac{2j'_1+1}{k'+2})\gamma(\frac{2j'_2+1}{k'+2})
\gamma(\frac{2j'_3+1}{k'+2})}}  \, 
\frac{P(j'_1+j'_2+j'_3+1)\, P(j'_{12})\, P(j'_{23})\, P(j'_{31})}{
P(2j'_1)\, P(2j'_2)\, P(2j'_3)}
\end{align}
and
\begin{align}
P(j')=\prod_{m=1}^{j'} \gamma({\textstyle\frac{m}{k'+2}})  \ ,\qquad
P(0)=1 \ ,\qquad
\gamma(x)=\frac{\Gamma(x)}{\Gamma(1-x)} \ .
\end{align}
The functions $P(j)$ are nonvanishing for $0 \leq j' \leq k'+1$. 
Therefore, $C'_{j'_1,j'_2,j'_3} \neq 0$, if 
\begin{align} \label{condjprime}
j'_1+j'_2+j'_3 \leq k' \ .
\end{align}

\noindent In $m'$-space the three-point function can be written as 
\begin{align} \label{3ptmspacesu2}
&\langle \Phi'_{j'_1,m'_1,\bar m'_1}(z_1,\bar z_1)
\Phi'_{j'_2,m'_2,\bar m'_2}(z_2,\bar z_2)
\Phi'_{j'_3,m'_3,\bar m'_3} (z_3, \bar z_3)\rangle
= \delta^2({\textstyle \sum_{a=1}^3 m'_a}) \,
{\cal D}^{j'_3}_{j'_1j'_2} \, 
\prod_{i<j} \frac{1} {|z_{ij}|^{2\Delta'_{ij}}} \ ,
\end{align}
where the $SU(2)$ operator algebra coefficients 
${\cal  D}^{j'_3}_{j'_1j'_2}$ can be found in \cite{Dotsenko} for the
case that
\begin{align}
m'_1=\bar m'_1 =j'_1-d \ ,\qquad 
m'_2=\bar m'_2 =j'_2 \ ,\qquad 
m'_3=\bar m'_3 =-j'_3=-(j'_1+j'_2-d) \ .
\end{align} 
They are given by (see Eq.~(2.46) in \cite{Dotsenko})
\begin{align}
{\cal D}^{j'_3}_{j'_1j'_2} = 
\frac{(2j'_2)!\, j'_{13}!}{j'_{12}!\, (2j'_3)!} \,
\prod^d_{i=1} 
\frac{ \gamma({\textstyle\frac{i}{k'+2}}) \,
\gamma(\frac{2j'_1+1+i}{k'+2})} \,
{\gamma(1+\frac{2j'_2-i+1}{k'+2})\,
\gamma(\frac{2j'_3-i+1}{k'+2})}
\sqrt{(a_{j'_1}')^{-1} \, a_{j'_2}' \, a_{j'_3}'}
\end{align}
with
\begin{align}
 a_{j'}'= \prod_{i=1}^{2j'}
 \frac{\gamma(1+\frac{i}{k'+2})}{\gamma(\frac{1+i}{k'+2})}
\,
\end{align}
and $d=j'_{12}=j'_1+j'_2-j'_3$. Here we performed a cyclic rotation of
the indices, $j'_1 \rightarrow j'_2$, {\it etc.} Of course, the
coefficients ${\cal D}^{j'_3}_{j'_1j'_2}$ are related to the structure
constants $C'_{j'_1,j'_2,j'_3}$ given by Eq.~(\ref{CSU2}). After some
algebra, one finds
\begin{align}
{\cal D}^{j'_3}_{j'_1j'_2} = 
\frac{\Gamma(j'_{13}+1)\, \Gamma(j'_{12}+1)}
{\Gamma(2j'_1+1)}\, C'_{j'_1,j'_2,j'_3} \equiv
\hat W(j'_i,m_i')\, C'_{j'_1,j'_2,j'_3}  \ .
\end{align}

\section{The correlators $G_3^{AA}$, $G_3^{AB}$, $G_3^{BA}$}
\label{AppB}

In this section we give some further details on the computation of the
terms $G_3^{AA}$, $G_3^{AB}$ and $G_3^{BA}$ appearing in the
worldsheet three-point function $G_3$ in section~\ref{sec3pt}. These 
terms can be written more explicitly by substituting the operators
${\cal W}_{j',m',m}$ and ${\cal W}^{0}_{j',m',m}$, as defined in
Eqns.~(\ref{W}) and (\ref{W0explicit}), into Eq.~(\ref{G3terms}).

\noindent For the term $G_3^{AA}$, we then get
\begin{align} \label{G3AA}
G^{AA}_3 &=g_s^{-2} 
b_{a_1} b_{a_2} b_{a_3} b_{\bar a_1} b_{\bar a_2} b_{\bar a_3} 
\langle  \Phi'_{j'_1,m'_1, \bar m'_1}\Phi'_{j'_2,m'_2,\bar m'_2}
\Phi'_{j'_3, m'_3,\bar  m'_3} 
\rangle \,{|z_{12}|^{-2}}
\\
&\quad\times \langle 
:\psi^{a_1}\bar \psi^{\bar a_1} \Phi_{j_1,m_1-a_1, \bar m_1-\bar a_1}:
:\psi^{a_2}\bar \psi^{\bar a_2} \Phi_{j_2,m_2-a_2,\bar m_2-\bar a_2}:
:J^{a_3}\bar J^{\bar a_3}\Phi_{j_3,m_3-a_3,\bar m_3-\bar a_3}: \rangle \,,
\nonumber
\end{align}
where we have to sum over $a_i$ and $\bar a_i$ ($i=1,2,3$). The factor
$|z_{12}|^{-2}$ is again the contribution from the
ghosts. Using Wick contraction and the OPE's~(\ref{currents}),
$G^{AA}_3$ can be rewritten as
\begin{align}
G^{AA}_3 &=g_s^{-2} 
b_{a_1} b_{a_2} b_{a_3} b_{\bar a_1} b_{\bar a_2} b_{\bar a_3} 
\langle  \Phi'_{j'_1,m'_1, \bar m'_1}\Phi'_{j'_2,m'_2,\bar m'_2}
\Phi'_{j'_3, m'_3,\bar  m'_3} 
\rangle 
  \frac{k^2}{4}\frac{\eta^{a_1a_2} 
  \eta^{\bar a_1 \bar a_2} }{|z_{13}|^2 |z_{23}|^2 |z_{12}|^{2}}
\nonumber\\  
  &\hspace{0.9cm}\times \left( 
     f_{a_3}^{m_1-a_1} f_{\bar a_3}^{\bar m_1-\bar a_1}
     \langle  \Phi_{j_1,m_1-a_1+a_3, \bar  m_1-\bar a_1+\bar a_3}
     \Phi_{j_2,m_2-a_2,\bar m_2-\bar a_2}
     \Phi_{j_3,m_3-a_3,\bar m_3-\bar a_3} \rangle \right. \nonumber\\
   &\hspace{1.1cm}
   + f_{a_3}^{m_2-a_2} f_{\bar a_3}^{\bar m_2-\bar a_2}
     \langle  \Phi_{j_1,m_1-a_1, \bar  m_1-\bar a_1}
     \Phi_{j_2,m_2-a_2+a_3,\bar m_2-\bar a_2+\bar a_3}
     \Phi_{j_3,m_3-a_3,\bar m_3-\bar a_3} \rangle\nonumber\\
   &\hspace{1.1cm}
    + f_{a_3}^{m_1-a_1} f_{\bar a_3}^{\bar m_2-\bar a_2}  
     \langle \Phi_{j_1,m_1-a_1+a_3, \bar  m_1-\bar a_1}
     \Phi_{j_2,m_2-a_2,\bar m_2-\bar a_2+\bar a_3}
     \Phi_{j_3,m_3-a_3,\bar m_3-\bar a_3}\rangle \nonumber\\
   &\hspace{1.1cm}
     \left.+ f_{a_3}^{m_2-a_2} f_{\bar a_3}^{\bar m_1-\bar a_1}
     \langle  \Phi_{j_1,m_1-a_1, \bar  m_1-\bar a_1+\bar a_3}
     \Phi_{j_2,m_2-a_2+a_3,\bar m_2-\bar a_2}
     \Phi_{j_3,m_3-a_3,\bar m_3-\bar a_3} \rangle \right) \ ,
\end{align}
with $f^m_{a}= (m-j+1, m, m+j-1)$ for ($a=+,0,-$). Due to the constraints 
$a_1 = -a_2$ and $\bar a_1 = - \bar a_2$, we effectively
sum only over four different indices ($a_1, a_3, \bar a_1,\bar a_3$
say). We therefore get $3^4=81$ terms which we sum up in the same way
as explained in detail for $G_3^{BB}$ in section~\ref{sec3pt}. We find
\begin{align}
G^{AA}_3 = g_s^{-2} g^{AA}(j'_i) \,C_{j_1, j_2, j_3}C'_{j'_1, j'_2, j'_3}
W(j_i-1;m_i) \hat W(j'_i;m'_i) 
\prod_{i<j}\frac{1}{|z_{ij}|^{2(\Delta_{ij}+\Delta'_{ij}+1)}} \ ,
\end{align}
with 
\begin{align} \label{gAA}
g^{AA}(j'_i) & = \frac{k^2}{16} \, (1+j'_1+j'_2-j'_3)^2 \ . 
\end{align}
Next, we consider the correlator $G^{AB}_3$ given by 
\begin{align} \label{G3AB}
G^{AB}_3 &=g_s^{-2} \frac{2}{k}
b_{a_1} b_{a_2} b_{a_3} b_{\bar a_1} b_{\bar a_2} A_{\bar a_3\bar b_3} 
\langle  \Phi'_{j'_1,m'_1, \bar m'_1}\Phi'_{j'_2,m'_2,\bar m'_2}
\Phi'_{j'_3, m'_3,\bar  m'_3} 
\rangle \langle  \bar \psi^{\bar a_1} \bar \psi^{\bar a_2} 
\bar \psi^{\bar a_3} \bar \psi^{\bar b_3} \rangle
\,|z_{12}|^{- 2}\\
&\quad\times \langle
:\psi^{a_1} \Phi_{j_1,m_1-a_1, \bar m_1-\bar a_1}:
:\psi^{a_2} \Phi_{j_2,m_2-a_2,\bar m_2-\bar a_2}:
: J^{a_3} \Phi_{j_3,m_3-a_3, \bar m_3-\bar a_3-\bar b_3}  : \rangle \ .
\nonumber
\end{align}
With the help of the OPE's~(\ref{currents}) and Eq.~(\ref{3ptpsi}), we
get 
\begin{align}
G^{AB}_3 &=g_s^{-2} 
b_{a_1} b_{a_2} b_{a_3} b_{\bar a_1} b_{\bar a_2} A_{\bar a_3\bar b_3} 
\langle  \Phi'_{j'_1,m'_1, \bar m'_1}\Phi'_{j'_2,m'_2,\bar m'_2}
\Phi'_{j'_3, m'_3,\bar  m'_3} 
\rangle  {k^2}\frac{\eta^{\bar a_1[\bar b_3} 
  \eta^{\bar a_3] \bar a_2} }{|z_{13}|^2 |z_{23}|^2 |z_{12}|^{2}}  
  \eta^{a_1a_2}\nonumber\\
  &\quad\times \left( 
     f_{a_3}^{m_1-a_1} 
     \langle  \Phi_{j_1,m_1-a_1+a_3, \bar  m_1-\bar a_1}
     \Phi_{j_2,m_2-a_2,\bar m_2-\bar a_2}
     \Phi_{j_3,m_3-a_3,\bar m_3-\bar a_3-\bar b_3} \rangle
            \right. \nonumber\\ 
   &\quad\hspace{0.2cm}
   \left.+ \,f_{a_3}^{m_2-a_2} 
     \langle  \Phi_{j_1,m_1-a_1, \bar  m_1-\bar a_1}
     \Phi_{j_2,m_2-a_2+a_3,\bar m_2-\bar a_2}
     \Phi_{j_3,m_3-a_3,\bar m_3-\bar a_3- \bar b_3} \rangle\right)\,,
\end{align}
with $f^m_{a}= (m-j, m, m+j)$ for ($a=+,0,-$), as before. Here we have
to sum over the seven indices $a_i, \bar a_i,$ and $b_3$
($i=1,2,3$). Again, due to several constraints,
\begin{align}
a_1 &= - a_2\,, \\
(\bar a_1 = -\bar b_3 \ ,\quad \bar a_2=-\bar a_3 \ ,\quad 
\bar a_3 \neq \bar b_3)\qquad 
&\hbox{or} \qquad (\bar a_1 = -\bar a_3 \ ,\quad \bar a_2=-\bar b_3 \ ,
\quad \bar a_3 \neq \bar b_3) \ , \nonumber
\end{align}
there are only 108 nonvanishing terms (out of $3^7$). The summation of
these terms yields 
\begin{align}
G^{AB}_3 = g_s^{-2} g^{AB}(j'_i) \,C_{j_1, j_2, j_3}C'_{j'_1, j'_2, j'_3}
W(j_i-1;m_i) \hat W(j'_i;m'_i) 
\prod_{i<j}\frac{1}{|z_{ij}|^{2(\Delta_{ij}+\Delta'_{ij}+1)}} \ ,
\end{align}
with 
\begin{align} \label{gAB}
g^{AB}(j'_i) & = \frac{k^2}{16} \, {(1+j'_1+j'_2-j'_3) 2 j'_3} \ . 
\end{align}
The same result is obtained for $G^{BA}_3$
({\it i.e.}\ $G^{BA}_3=G^{AB}_3$).

\section{Product of  $SL(2)_k$ and $SU(2)_{k'}$ structure constants} 
\label{appendixproduct}

In this appendix we compute the product of the $SL(2)_k$ and
$SU(2)_{k'}$ structure constants. Here we work with the supersymmetric
levels, {\it i.e.}\ we need to shift $k \rightarrow k+2$ and 
$k'\rightarrow k'-2$ in $C_{j_1,j_2,j_3}$ and $C'_{j'_1,j'_2,j'_3}$,
respectively, and identify $k=k'$.  From Eq.~(\ref{Gplus1}) we then
find a simple relation between the functions $G(j)$ and $P(j')$
appearing in the $SL(2)$ and $SU(2)$ structure constants, 
\begin{align}
G(-j)= \frac{1}{\gamma(\frac{j-1}{k})} 
\cdots\frac{1}{\gamma(\frac{1}{k})}G(-1)= \frac{G(-1)}{P(j-1)} \ .
\end{align}
Substituting this into the definition of the $SL(2)$ structure
constant $C_{j_1,j_2,j_3}$ yields
\begin{align} 
C_{j_1,j_2,j_3} &=  \frac{1}{(2\pi)^2}\, 
\frac{1}{2\pi^3} \frac{k\, P(2j_1-2)P(2j_2-2)P(2j_3-2)}
{ \nu^{j_1+j_2+j_3-2} P(j_1+j_2+j_3-2) P(j_{12}-1) P(j_{23}-1)
P(j_{31}-1) }\, \nonumber\\
&= \frac{1}{(2\pi)^3}\sqrt{\frac{\gamma({\textstyle\frac{1}{k}})}{
\gamma(\frac{2j_1-1}{k})\gamma(\frac{2j_2-1}{k})
\gamma(\frac{2j_3-1}{k})}} \frac{1}{C'_{j'_1,j'_2,j'_3}}
\frac{k}{\pi^2 \nu^{j_1+j_2+j_3-2}} \ ,
\end{align}
where $j'_i = j_i-1$.  Using the definitions of $B(j)$ and $\nu$,
Eq.~(\ref{Bj}), we finally get
\begin{align} \label{magic}
C_{j_1,j_2,j_3}C'_{j'_1,j'_2,j'_3}
= \sqrt{ B(j_1)B(j_2)B(j_3)} \ .
\end{align}
The (trivial) numerical coefficient in (\ref{magic}) is a consequence
of our choice of normalisation for the $SL(2)$ amplitudes --- see 
footnote~9.

\end{document}